# QUANTILE-STRATIFIED SAMPLING FOR MULTIVARIATE NORMAL SIMULATIONS AND OTHER MULTIVARIATE DISTRIBUTIONS

B. O'NEILL,[*] *ACIL Allen*[**]

**Abstract**

In this paper we show how to extend quantile-stratified sampling to produce simulations from various multivariate distributions. These simulations have desirable space-filling and coverage properties relative to simulation using IID sampling. We examine the coverage performance of these simulations against IID sampling by looking at plots of ordered log-density values from the simulations.

## 1. Introduction

Quantile-stratified (QS) sampling is a method for generating pseudo-random values from a known probability distribution by splitting the support of the distribution into equiprobable "quantile-blocks" and then generating one value from each of these intervals. The quantile-blocks are formed as contiguous intervals that partition the support of the distribution, with the boundaries of these blocks being the quantiles of the distribution. This method is described for one-dimensional probability distributions in O'Neill (2025).

In contrast to IID sampling, QS sampling ensures greater consistency in the spread of values over the support of the distribution, which gives lower variance to the order statistics of the generated sample. These methods differ in a manner that is analogous to the difference between simple-random-sampling with replacement and simple-random-sampling without replacement. In the case of IID samples the values are effectively generated with replacement over all the quantile-blocks and in the case of QS sampling the values are generated without replacement over the quantile-blocks, with one value occurring in each block.

The QS sampling method is a "space filling" design (in the language of sampling design theory) in the sense that QS sampling has a greater tendency than IID sampling to cover contiguous areas in the support of the marginal sampling distribution that the user is sampling from (see e.g., Santner, Williams and Notz 2003, Ch 5; Draguljic, Santner and Dean 2012). The most widely used method of producing a space-filling design is the Latin hypercube (see e.g., Street and Street 1987; Tang 1993; Cox and Reid 2000, p. 181-182; Hinkelmann 2011, 401-419; Iman 2013), and univariate QS sampling is equivalent to one-dimensional Latin hypercube sampling. Consequently, the general method of Latin hypercube sampling is one possible generalisation of QS sampling to multivariate distributions, but it is possible to construct alternative methods that decompose a multivariate space in a different way to the Latin hypercube method. In particular, one possible generalisation is to decompose the support of a multivariate distribution into quantile-blocks corresponding to contiguous ranges of the output of the density function (equivalently the log-density) so that the selected multivariate quantile-blocks "ripple outward" from the mode of the distribution with boundaries given by the contours of the density.

[*] E-mail address: ben.oneill@hotmail.com.
[**] Level 6, 54 Marcus Clarke Street, Canberra ACT 2601, Australia.

In the case of a univariate distribution, QS sampling can be characterised as follows. Suppose we wish to generate $n$ pseudo-random values from a distribution with known density function $f$, distribution function $F$ and quantile function $Q$. We generate a QS sample by generating simple random samples without replacement over the indices for quantile-blocks and then sampling over the conditional distribution over each of the quantile-blocks:

$$S_1^*, \dots, S_n^* \sim \text{SRSWOR}\{1, \dots, n\} \qquad U_i^* \sim \text{U}\left(\frac{S_i^* - 1}{n}, \frac{S_i^*}{n}\right) \qquad X_i^* \equiv Q(U_i^*).$$

Throughout the remainder of the paper we refer to an IID sample and a QS sample (the latter generated by the process described above) using the following shorthand:

$$X_1, \dots, X_n \sim \text{IID } f,$$

$$X_1^*, \dots, X_n^* \sim \text{QS } f.$$

The properties of QS sampling from a univariate distribution have been examined in detail and contrasted with IID sampling in O'Neill (2025). That analysis showed that QS sampling gives greater consistency in order statistics and greater accuracy when used in simulation-based estimation methods such as importance sampling. The goal of the present paper is to extend this analysis to show how QS sampling can be extended to multivariate settings by constructing quantile-blocks bounded by density (equivalently log-density) contours. To build up this method in a clear way, we will implement QS sampling for a multivariate normal distribution, then broader multivariate elliptically contoured distributions, and then other multivariate distributions. We will show how multivariate QS sampling can be implemented using exact computational methods, and also how it can be approximated with quasi-QS sampling.

## 2. QS sampling from the multivariate normal distribution

Before showing how to generate a QS sample from the multivariate normal distribution, we first remind ourselves of how to generate a single random vector from this distribution using the Box-Muller transform (Ross 2002, pp. 279-281). Suppose we want to generate a random vector $\boldsymbol{Y} \sim \text{N}(\boldsymbol{\mu}, \boldsymbol{\Sigma})$ with vector length $k$ and let $\boldsymbol{\Sigma}^{1/2}$ be the principal square root of the variance matrix. We can generate this random vector using the Box-Muller transformation as:

$$\boldsymbol{Y} = \boldsymbol{\mu} + R \cdot \frac{\boldsymbol{\Sigma}^{1/2}\boldsymbol{Z}}{||\boldsymbol{Z}||} \qquad R \sim \text{Chi}(k) \qquad \boldsymbol{Z} \sim \text{N}(\boldsymbol{0}, \mathbf{I}).$$

For the purposes of generating a QS sample the value of this method is that it takes advantage of the spherical symmetry of the normal distribution. The elements $R$ and $\boldsymbol{Z}$ are independent, and moreover, the radius $R$ is a monotone function of the normal density (the larger the radius the lower the density at that point) whereas the direction $\boldsymbol{Z}$ is a constant function of the normal density (i.e., the density at the point is the same regardless of the direction). This means that there is a "natural" ordering to form quantile-blocks based on the radius $R$, which is proportionate to the Mahalanobis distance.[1] The quantile-blocks are regions bounded by ellipses emanating out from the mean and giving the level sets for the multivariate normal density.

To generate a QS sample of $n$ values from the **multivariate normal distribution** we proceed as follows. Let $f_R$ and $Q_R$ be the density and quantile functions for the chi distribution with $k$ degrees-of-freedom and generate the QS sample $R_1^*, \ldots, R_n^* \sim \text{QS } f_R$ as follows:

$$S_1^*, \ldots, S_n^* \sim \text{ SRSWOR}\{1, \ldots, n\} \qquad U_i^* \sim \text{U}\left(\frac{S_i^* - 1}{n}, \frac{S_i^*}{n}\right) \qquad R_i^* \equiv Q_R(U_i^*).$$

We then obtain a QS sample from the multivariate normal distribution by taking:

$$\boldsymbol{Y}_i^* = \boldsymbol{\mu} + R_i^* \cdot \frac{\boldsymbol{\Sigma}^{1/2}\boldsymbol{Z}_i}{||\boldsymbol{Z}_i||} \qquad \boldsymbol{Z}_i \sim \text{N}(\boldsymbol{0}, \mathbf{I}).$$

This method yields a QS sample from the multivariate normal distribution with mean vector $\boldsymbol{\mu}$ and variance matrix $\boldsymbol{\Sigma}$. Because the quantile-blocks are regions bonded by ellipses giving the level sets of the multivariate normal density, the resulting QS sample is guaranteed to include points spread consistently over different levels of the Mahalanobis distance. As in QS sampling from a univariate distribution, this gives greater consistency in coverage.

[1] The Mahalanobis distance for the random vector $\boldsymbol{Y}$ is obtained by dividing the radius by $\sqrt{k}$.

O'Neill (2025) establishes that the univariate QS sampling method generates values from the marginal distribution of interest, which ensures that we have $R_i^* \sim \text{Chi}(k)$, so the Box-Muller transformation then ensures that $\boldsymbol{Y}_i^* \sim \text{N}(\boldsymbol{\mu}, \boldsymbol{\Sigma})$. The values generated in the QS sample are not independent owing to the sampling without replacement from the quantile-blocks. Compared to an IID sample from the multivariate normal distribution, the QS sample typically exhibits greater regularity in relation to the Mahalanobis distances of the simulated values. This greater regularity manifests as lower variance in the order statistics and negative correlation between the Mahalanobis values (see O'Neill 2025 for details, esp. Theorems 3-4).

In Figure 1 below we show samples of $n = 30$ values generated from the standard bivariate normal distribution for QS sampling and IID sampling. The difference between these methods is not obvious from the plot, as they both appear to give random coverage as you would expect for this distribution. Though imperceptible in this plot, the difference is that the QS sample uses sampling without replacement to give one value in each of the quantile-blocks, which in the case of the standard bivariate normal are circular rings emanating out from the origin. The absence of any difference that is perceptible in the scatter plot is a salient feature of the QS sample, since it means that the greater regularity in the sample does not come at the expense of any appearance of randomness relative to IID sampling.

In Figure 2 we show the ordered log-density values of the points in the same samples. Here it becomes clear that the QS sample has greater regularity over the range of log-density values, owing to its greater regularity across the quantile-blocks. As can be seen from the plot, the QS sample has a much smoother transition of the ordered log-density values in the sample. This is because the partition into quantile-blocks also implicitly partitions the corresponding range of possible values of the log-density. Since each value in the sample is restricted to a unique quantile-block it is also restricted to a unique range of possible log-density values giving rise to the exhibited smoothness. In the case of the multivariate normal distribution the log-density curve is effectively the same as the log-density curve for the chi distribution, owing to the spherical symmetry of the distribution and the rate at which the density decays from its mode.

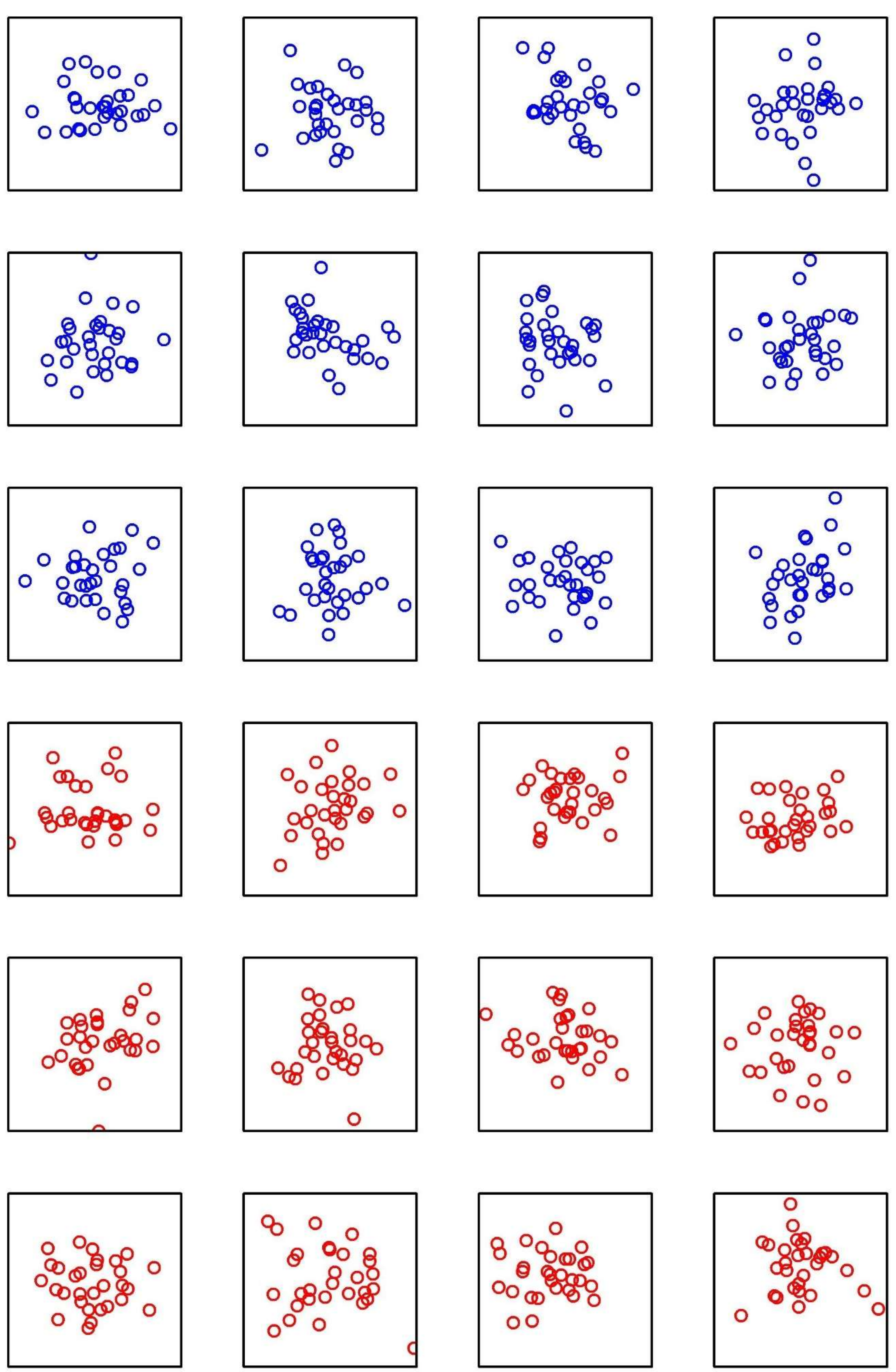

**FIGURE 1:** Plots of ordered samples of $n = 30$ data points from the standard bivariate normal distribution (QS sampling in blue — IID sampling in red))

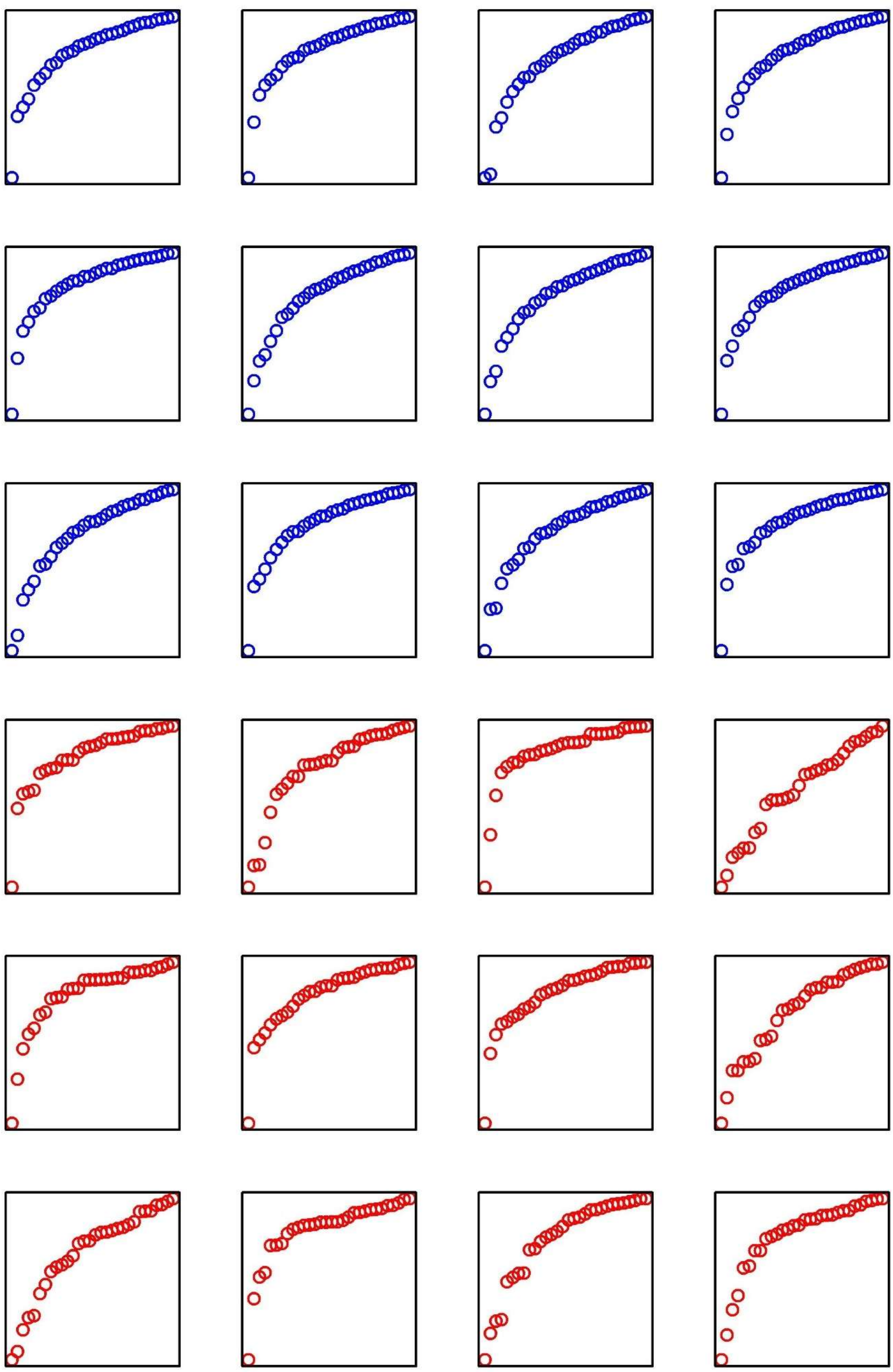

**FIGURE 2:** Plots of ordered log-density values for bivariate normal samples
(QS sampling in blue — IID sampling in red))

## 3. QS sampling from multivariate elliptically contoured distributions

Generating a QS sample for the multivariate normal distribution is relatively simple due to the fact that there are simple formulae and methods to simulate the distance from the mean to the level sets, and to simulate a random point on the level sets. This allowed us to use QS sampling for the univariate distance measure, which gave a QS sample for the multivariate distribution. Another broad class of cases that form a relatively simple extension of the multivariate normal case, with a similar decomposition, is the class of all **elliptically contoured distributions** (see e.g., Chmielewski 1981; Johnson 1987, pp. 106-124). These are multivariate distributions that have a density kernel of the form:

$$f_Y(\boldsymbol{y}|\boldsymbol{\mu}, \boldsymbol{\Sigma}, g) \propto g((\boldsymbol{y}-\boldsymbol{\mu})^{\text{T}}\boldsymbol{\Sigma}^{-1}(\boldsymbol{y}-\boldsymbol{\mu})) \qquad \boldsymbol{y} \in \mathbb{R}^k,$$

where $g: \mathbb{R}_{0+} \to \mathbb{R}_{0+}$ is a function operating on the quadratic form that decreases sufficiently fast it its tail to give a proper density. In the case where $g$ is a monotone decreasing function the density is strictly quasi-concave with mode at $\boldsymbol{\mu}$. The normal distribution occurs in the case where $g(r^2) = \exp(-r^2/2)$.

For all these distributions, a natural candidate for the distance measure is the quadratic form:

$$R = \sqrt{(\boldsymbol{Y}-\boldsymbol{\mu})^{\text{T}}\boldsymbol{\Sigma}^{-1}(\boldsymbol{Y}-\boldsymbol{\mu})},$$

Let $\boldsymbol{Y} \sim \text{EC}_g(\boldsymbol{\mu}, \boldsymbol{\Sigma})$ with the density kernel shown above. Then the density kernel for $R$ is:

$$f_R(r|g) \propto r^{k-1} g(r^2) \qquad r \geq 0.$$

With a bit of computational work, the density kernel can be used to find the other probability functions for $R$. Letting $f_R$ and $Q_R$ be the density and quantile functions for this distribution, we can generate the QS sample $R_1^*, \dots, R_n^* \sim \text{QS}\ f_R$ as follows:

$$S_1^*, \dots, S_n^* \sim \text{SRSWOR}\{1, \dots, n\} \qquad U_i^* \sim \text{U}\left(\frac{S_i^*-1}{n}, \frac{S_i^*}{n}\right) \qquad R_i^* \equiv Q_R(U_i^*).$$

We then obtain a QS sample from the multivariate elliptically contoured distribution by taking:

$$\boldsymbol{Y}_i^* = \boldsymbol{\mu} + R_i^* \cdot \frac{\boldsymbol{\Sigma}^{1/2}\boldsymbol{Z}_i}{||\boldsymbol{Z}_i||} \qquad \boldsymbol{Z}_i \sim \text{N}(\boldsymbol{0}, \mathbf{I}).$$

This method yields a QS sample $\boldsymbol{Y}_1^*, \dots, \boldsymbol{Y}_n^* \sim \text{QS}\ \text{EC}_g(\boldsymbol{\mu}, \boldsymbol{\Sigma})$. Just as for the multivariate normal distribution, the level sets of the distribution are elliptical contours and so the quantile-blocks for the simulation are regions bounded on both sides by those ellipses. By forcing each of the quantile-blocks to contain exactly one simulated value, the QS sample gives greater regularity in the ordered log-density values for the simulated points than an IID sample.

There are several elliptically contoured distributions in common use in statistical analysis. One of these is the **multivariate T-distribution**, which generalises the normal distribution by adding a degrees-of-freedom parameter to represent "fatter tails" than the normal. For the multivariate T-distribution with dimension $k$ and $v$ degrees-of-freedom we have:

$$g(r^2) = \left(1 + \frac{r^2}{v}\right)^{-\frac{v+k}{2}},$$

which gives the density kernel for the distance measure as:

$$f_R(r|k,v) \propto r^{k-1}\left(1 + \frac{r^2}{v}\right)^{-\frac{v+k}{2}} \qquad r \geq 0.$$

We can obtain the corresponding distribution function by integrating the density kernel. Using the change of variable $s = r^2/v$ we have the differential relationship $ds = (2r/v)\,dr$ and the integral can be written as:

$$\begin{aligned}
\text{INT}(t,k,v) &\equiv \int_0^t r^{k-1}\left(1 + \frac{r^2}{v}\right)^{-\frac{v+k}{2}} dr \\
&= \frac{v^{k/2}}{2}\int_0^{t^2/v} s^{(k-2)/2}(1+s)^{-(v+k)/2}\, ds \\
&= \frac{v^{k/2}}{2}\int_0^{t^2/v} \frac{s^{k/2-1}}{(1+s)^{k/2+v/2}}\, ds \\
&= \frac{v^{k/2}}{2}\cdot \text{IBU}\left(\frac{v}{v+t^2}\,\middle|\,\frac{k}{2},\frac{v}{2}\right) \\
&= \frac{v^{k/2}}{2}\cdot \text{B}\left(\frac{k}{2},\frac{v}{2}\right)\cdot \text{RIBU}\left(\frac{v}{v+t^2}\,\middle|\,\frac{k}{2},\frac{v}{2}\right),
\end{aligned}$$

where B is the beta function, IBU is the upper incomplete beta function, and RIBU is the upper regularised incomplete beta function. Using this integral result, the cumulative distribution function for the distance measure $R$ is:

$$F_R(r|k,v) = \text{RIBU}\left(\frac{v}{v+r^2}\,\middle|\,\frac{k}{2},\frac{v}{2}\right) \qquad r \geq 0.$$

Setting $p = F_R(Q_R(p)|k,v)$ and solving then yields the corresponding quantile function for the distance measure $R$, which is:

$$Q_R(p) = \sqrt{v \cdot \frac{1 - \text{RIBU}^{-1}\left(p\,\middle|\,\frac{k}{2},\frac{v}{2}\right)}{\text{RIBU}^{-1}\left(p\,\middle|\,\frac{k}{2},\frac{v}{2}\right)}} \qquad 0 \leq p \leq 1.$$

Another elliptically contoured distribution in reasonably common use in statistical analysis is the **multivariate logistic distribution**, which generalises the logistic distribution to multiple dimensions. For the multivariate logistic distribution with dimension $k$ we have:

$$g(r^2) = \frac{\exp(r)}{(1+\exp(r))^2},$$

which gives the density kernel for the distance measure as:

$$f_R(r|k) \propto r^{k-1} \frac{\exp(r)}{(1+\exp(r))^2} \qquad r \geq 0.$$

Applying integration by parts[2] and then expanding the geometric series, we obtain:

$$\begin{aligned}
\text{INT}(t,k) &\equiv \int_0^t r^{k-1} \frac{\exp(r)}{(1+\exp(r))^2} dr \\
&= \frac{t^{k-1}}{1+\exp(t)} + (k-1) \int_0^t \frac{r^{k-2}}{1+\exp(r)} dr \\
&= \frac{t^{k-1}}{1+\exp(t)} + (k-1) \sum_{i=0}^{\infty} (-1)^i \int_0^t r^{k-2} \exp(-(i+1)r)\, dr \\
&= \frac{t^{k-1}}{1+\exp(t)} + (k-1) \sum_{i=0}^{\infty} \frac{(-1)^i}{(i+1)^{k-1}} \int_0^t s^{k-2} \exp(-s)\, ds \\
&= \frac{t^{k-1}}{1+\exp(t)} + (k-1)\gamma(k-1,t) \sum_{i=0}^{\infty} \frac{(-1)^i}{(i+1)^{k-1}} \\
&= \frac{t^{k-1}}{1+\exp(t)} + (k-1)\gamma(k-1,t)\, \eta(k-1) \\
&= \frac{t^{k-1}}{1+\exp(t)} + [\gamma(k,t) - t^{k-1}\exp(-t)]\, \eta(k-1) \\
&= \frac{t^{k-1}}{1+\exp(t)} + \Gamma(k)\, [F_{\text{Ga}}(t|k,1) - f_{\text{Ga}}(t|k,1)]\, \eta(k-1),
\end{aligned}$$

where $\gamma$ is the lower incomplete gamma function and $\eta$ is the Dirichlet eta function. In the special case $t \to \infty$ this simplifies to $\text{INT}(\infty,k) = \Gamma(k)\,\eta(k-1)$. Using this integral result, the cumulative distribution function for the distance measure $R$ is:

---

[2] The first step is the application of integration by parts, using the functions:

$$u(r) = r^{k-1} \qquad \acute{u}(r) = (k-1)r^{k-2},$$

$$v(r) = -\frac{1}{1+\exp(r)} \qquad \acute{v}(r) = \frac{\exp(r)}{(1+\exp(r))^2}.$$

$$F_R(r|k,v) = \frac{1}{\Gamma(k)\,\eta(k-1)} \cdot \frac{r^{k-1}}{1+\exp(r)} + [F_{\text{Ga}}(r|k,1) - f_{\text{Ga}}(r|k,1)] \qquad r \geq 0.$$

Setting $p = F_R(Q_R(p)|k,v)$ and solving then yields the corresponding quantile function for the distance measure $R$, which does not have a closed form expression or an expression in terms of standard special functions. The quantiles for QS sampling can be computed numerically for the procedure using this equation.

There are some other elliptically contoured distributions that are used occasionally in statistics, with higher levels of complexity in the computation of the resulting quantiles for the distance measure. Some examples of these distributions include the symmetric multivariate Laplace distribution, the symmetric multivariate general hyperbolic distribution, and the symmetric multivariate stable distribution. The last of these distributions is generally defined through its characteristic function and it does not even have an analytic expression for the density kernel, which presents further challenges for computing the relevant quantiles. In general, it may be challenging to compute the relevant quantiles for an arbitrary elliptically contoured distribution to implement the QS sampling method, but this is the only real challenge of the method, since the rest of the algorithm is straightforward.

## 4. Extension of QS sampling to other quasi-concave multivariate distributions

We have seen how to generate a QS sample for the normal distribution and other elliptically contoured distributions. This approach generalises (in principle) for any strictly quasi-concave distribution. In the case of the normal distribution we formulated the QS sampling scheme by using a scalar random variable $R$ to measure the "distance" from the mode as a correspondence to the level curves of the density, and we used an independent random vector to generate a random point over the level curve.[3] We extended this to elliptically contoured distributions using the sample method, with the only additional complexity being some potential additional difficulty computing the quantiles at issue. In principle, this method can be used for any strictly quasi-concave density, but it raises trickier computational/simulation requirements. We will illustrate the method here for a strictly quasi-concave continuous multivariate distribution.

---

[3] In the case of the normal distribution, we used the random vector $\mathbf{\Sigma}^{1/2}\mathbf{Z}/||\mathbf{Z}||$ to give a random point over the surface of an ellipse and then we scaled by the distance and centered on the mean vector to get a final point on the level curve.

Suppose that our goal is to generate some $n$-dimensional random vector $\boldsymbol{Y}$ which follows some **strictly quasi-concave continuous multivariate distribution** with unique mode $\boldsymbol{y}_{\text{mode}}$. To do this, we can formulate a scalar random variable $R$ that is a measure of "distance" from the mode, corresponding to the level curves of the density. It is natural to set this measure so that it is zero at the mode and increases continuously as the log-density of the distribution decreases, using a one-to-one continuous mapping to the log-density. For a log-density distance $\ell \in \mathbb{R}$, we take a corresponding value $R = r$ defined by a function $T: \mathbb{R} \rightarrow \mathbb{R}_{0+}$ of the following form:

$$R = T(\ell) \qquad T(0) = 0 \qquad T'(\ell) > 0.$$

Let $f_{\boldsymbol{Y}}$ denote the density function for $\boldsymbol{Y}$ and define the level curve mapping:

$$\boldsymbol{\mathcal{Y}}(r) \equiv \{\boldsymbol{y} \in \mathbb{R}^n \mid \log f_{\boldsymbol{Y}}(\boldsymbol{y}_{\text{mode}}) - \log f_{\boldsymbol{Y}}(\boldsymbol{y}) = T^{-1}(r)\} \qquad r \geq 0.$$

We can see that $\boldsymbol{\mathcal{Y}}(0) = \{\boldsymbol{y}_{\text{mode}}\}$ and as we increase $r$ the sets $\boldsymbol{\mathcal{Y}}(r)$ correspond to level curves for lower density values that are further away from the mode.

From the distribution of $\boldsymbol{Y}$ there is some corresponding distribution for $R$, with density that we will denote by $f_R$. Conditional on $R$ (i.e., conditional on the point being on a particular level curve) there is then some conditional distribution for $\boldsymbol{Y}$ over that level curve, with density that we will denote by $f_{\boldsymbol{Y}|R}$. This distance-based decomposition allows us to generate the random variable $\boldsymbol{Y}$ through the two-step process:

$$\boldsymbol{Y}|R \sim f_{\boldsymbol{Y}|R} \qquad R \sim f_R.$$

Turning this generation method into a QS sample is possible in principle, but may involve some computational challenges. We generate values of the distance value $R$ using QS sampling and then we generate values of $\boldsymbol{Y}$ over the conditional distribution on the relevant level curves. That is, we generate the QS sample $\boldsymbol{Y}_1^*, \dots, \boldsymbol{Y}_n^* \sim \text{QS } f_{\boldsymbol{Y}}$ using the following method:

$$\boldsymbol{Y}_i^*|R_i^* \sim f_{\boldsymbol{Y}|R} \qquad R_1^*, \dots, R_n^* \sim \text{QS } f_R.$$

Since the above generation method preserves the two-step method for generating $\boldsymbol{Y}$, it preserves the desired marginal distribution for sampling. By generating the distance values in the scheme using QS sampling, we ensure that we generate exactly one distance value $R_i^*$ in each quantile-block based on the quantiles of the log-density curve. This means that we ensure that we have one generated sample point $\boldsymbol{Y}_i^*$ in each quantile-block represented as a region bounded by the level curves of the density function.

As in the case of the multivariate normal distribution, it may be possible to facilitate sampling from the conditional distribution $f_{\boldsymbol{Y}|R}$ (i.e., the conditional distribution given that a point lies on a particular level curve) by using various intermediate random variables. For a value $R = r$ the conditional density has support on the level curve $\boldsymbol{\mathcal{Y}}(r)$. It may be possible to generate the value $\boldsymbol{Y}$ from this conditional distribution using an intermediate quantity $\boldsymbol{Z}$ via a transformation:

$$\boldsymbol{Y} = \boldsymbol{f}(R, \boldsymbol{Z}) \in \boldsymbol{\mathcal{Y}}(r) \qquad \boldsymbol{Z}|R \sim \boldsymbol{f}_{\boldsymbol{Z}|R}.$$

In some cases, it may even be possible to use an intermediate quantity $\boldsymbol{Z}$ that is independent of $R$, giving a form that is further simplified. (By way of reminder, in the case of the multivariate normal distribution, this was done using the Box-Muller transform.) Given an intermediate quantity allowing this type of transformation for the output $\boldsymbol{Y}$, we can generate the QR sample $\boldsymbol{Y}_1^*, \dots, \boldsymbol{Y}_n^* \sim \text{QS } f_{\boldsymbol{Y}}$ using the following elaborated method:

$$\boldsymbol{Y}_i^* = \boldsymbol{f}(R_i^*, \boldsymbol{Z}_i^*) \qquad \boldsymbol{Z}_i^*|R_i^* \sim \text{IID } \boldsymbol{f}_{\boldsymbol{Z}|R} \qquad R_1^*, \dots, R_n^* \sim \text{QS } f_R.$$

This general extension for QS sampling is possible in principle, but it gives rise to some thorny computational challenges in all but the most structured distributions. Firstly, for an arbitrary strictly quasi-concave continuous multivariate distribution, it may be difficult to compute the quantiles of the corresponding distribution for the distance measure. Secondly, even once these are computed, it may be hard to simulate from the conditional densities over the level curves for a fixed value of the distance measure. These may be significant computational challenges in their own right, and they may present barriers to implementation.

The above method allows the generation of QS samples for strictly quasi-concave multivariate distributions with a unique mode, but it can be extended to arbitrary distributions by relaxing the requirement for a unique mode and adapting the method to the possibility of multiple modes or a density that can approach infinity at some points (so that there is no mode). In principle, this extension can be done using a distance variable $R$ that measures distance based on relative density, relative to an arbitrary point with non-zero finite density, and which can take on both positive values (representing being further from the mode than that point) and negative values (representing being closer to the mode than that point). To do this, we pick an arbitrary point $\boldsymbol{y}_*$ with non-zero finite density and we set:

$$R = T\left(\log\left(\frac{f_{\boldsymbol{Y}}(\boldsymbol{y}_*)}{f_{\boldsymbol{Y}}(\boldsymbol{Y})}\right)\right) = T(\log f_{\boldsymbol{Y}}(\boldsymbol{y}_*) - \log f_{\boldsymbol{Y}}(\boldsymbol{Y})).$$

This extended version of the "distance" random variable can take both positive or negative values, with negative values occurring for points $\boldsymbol{y}$ with $f_{\boldsymbol{Y}}(\boldsymbol{y}) > f_{\boldsymbol{Y}}(\boldsymbol{y}_*)$. The remainder of the procedure is the same — we generate a QS sample $R_1^*, \dots, R_n^* \sim \text{QS } f_R$ and then generate the values for $\boldsymbol{Y}_i^* | R_i^* \sim f_{\boldsymbol{Y}|R}$ which are variables over the level sets of the density corresponding to the relevant distance value. Again, although this is possible in principle, there may be several computational challenges, including computation of the quantile function for the distance $R$ and simulation of resulting values over the level sets for a fixed value of the distance measure. These may be significant computational challenges that present barriers to implementation.

## 5. Quasi-QS sampling for arbitrary multivariate distributions

We have seen that there are some specific multivariate distributions where it is relatively simple to produce a QS sample in a natural way. Although it is possible in principle to extend these methods to arbitrary multivariate distributions, doing so involves significant computational challenges that may outweigh the value of the method. To avoid these issues, it is possible to fall back on generating a quasi-QS sample, as described in O'Neill (2025).

To implement this method, suppose that we are able to simulate values from the multivariate sampling distribution via some known simulation method (e.g., direct IID sampling from a known simulation method, or approximate sampling using an applicable MCMC method). It is not strictly necessary for the simulated values to be independent, but they should have low average dependence (e.g., from a long MCMC chain or similar method) and they should be marginally distributed according to the target multivariate distribution. (In the case of MCMC methods, this marginal result holds only approximately, but it can be made close by using a sufficient "burn-in" period.) Given a simulation method, suppose we set a multiplier $m \in \mathbb{N}$ and we generate values $\boldsymbol{y}_1, \dots, \boldsymbol{y}_{mn}$ from the multivariate distribution at issue, using the available sampling method. We then form the ranked statistics $\boldsymbol{y}_{[1]}, \dots, \boldsymbol{y}_{[nm]}$ using the ranking:

$$\log f_{\boldsymbol{Y}}(\boldsymbol{y}_{[i]}) \leq \log f_{\boldsymbol{Y}}(\boldsymbol{y}_{[i+1]}) \qquad \text{for all } i.$$

Finally, we generate the quasi-QS sample $\boldsymbol{Y}_1^*, \dots, \boldsymbol{Y}_n^*$ by taking:

$$\text{Quasi-QS:} \qquad \begin{matrix} T_1, \dots, T_n \sim \text{SRSWR}\{1, \dots, m\} \\ S_1^*, \dots, S_n^* \sim \text{SRSWOR}\{1, \dots, n\} \end{matrix} \qquad \boldsymbol{Y}_i^* = \boldsymbol{y}_{[(S_i^*-1)m+T_i]}.$$

An overview of the general method of quasi-QS sampling and its moment results is given in O'Neill (2025). The general idea of the method is that for a sufficiently large multiplier $m$, the statistics $\boldsymbol{y}_{[m]}, \boldsymbol{y}_{[2m]}, \dots, \boldsymbol{y}_{[nm]}$ will act as representative points that have a relative log density that approximates the corresponding quantiles $1/n, 2/n, \dots, 1$ for the log-density ratio $R$. That is, if we take $Q_R$ to be the true quantile function for $R$ then we have the approximation:

$$\log\left(\frac{f_Y(\boldsymbol{y}_*)}{f_Y(\boldsymbol{y}_{[sm]})}\right) \approx T^{-1}(Q_R(s/n)) \qquad \text{for all } s = 1, \dots, n.$$

This means that taking a random selection from the values $\boldsymbol{y}_{[(s-1)m+1]}, \dots, \boldsymbol{y}_{[sm]}$ approximates taking a random sample from the true quantile block:

$$\boldsymbol{\mathcal{Y}}_s \equiv \bigcup \{\boldsymbol{\mathcal{Y}}(r) | Q_R(s-1/n) < r \leq Q_R(s/n)\}.$$

The advantage of quasi-QS sampling is that is only requires a simulation method for generating values from the multivariate distribution at issue (or an approximate method such as MCMC) combined with the ability to compute and order the log-density at the simulated points. This reduces the computational difficulty of the method significantly and makes it simple to apply to an underlying base sample generated from the distribution. Of course, the drawback of the method is that it requires a multiplier for the number of points generated, and since we want the multiplier to be high enough to give a good approximation to the true quantile blocks, the generation of a quasi-QS sample will usually involve generation of an underlying base sample that is a substantial multiplier higher than the desired sample size. This means that quasi-QS sample (with a reasonably high multiplier) is not computationally efficient compared to IID sampling or other methods that produce the base sample.

**EXAMPLE:** Suppose we wish to generate a QS sample with $n = 30$ points from a three-element Dirichlet distribution with concentration parameter $\boldsymbol{\alpha} = (1.6, 1.3, 2.0)$. Obtaining an exact QS sample involves computational difficulties around simulation of the distance measure from the mode and corresponding simulation on level sets of the distribution. Suppose we decide to avoid these difficulties by instead generating a quasi-QS sample using a multiplier $m = 40$, taking an underlying sample of $nm = 1200$ IID points as the base of the quasi-QS sample. In Figure 3 below we show one simulation of a quasi-QS sample of this kind, compared to an IID sample of the same sample size. The quasi-QS sample is more "space filling" in the sense that it covers a greater variation of regions for the log-density of the distribution. This is reflected in the smoother log-density plot for the quasi-QS sample.

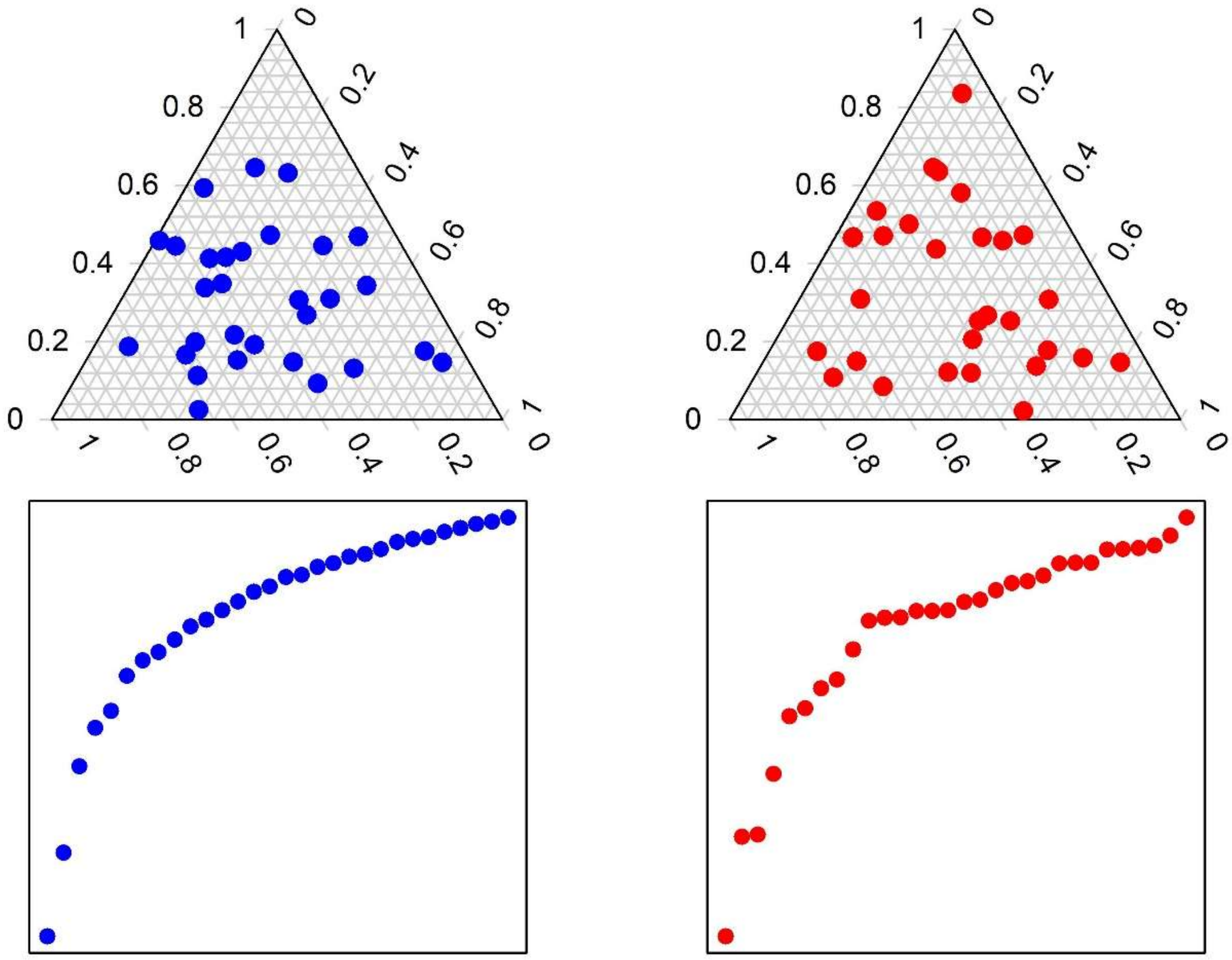

**FIGURE 3:** Ternary plots and ordered log-density plots for Dirichlet samples
(Ternary plot shows the three elements clockwise from the left)
(Quasi-QS sampling in blue — IID sampling in red)

As can be seen from the above example, it is quite simple to generate a quasi-QS sample for a multivariate distribution given some underlying simulation method allowing generation of a base sample. The quasi-QS sample can be made to be close to a true QS sample by setting a high multiplier $m$, but this makes the sampling method computationally inefficient compared to using the entire base sample. It is notable in the example above that the quasi-QS sample of $n = 30$ point was generated using an underlying IID base sample of $nm = 1200$ points (but our figure compares it to an IID sample with $n = 30$ points). Because of the computational burden of this method, quasi-QS sampling is not competitive as a substitute for direct sampling for estimation problems involving a simulated sample. Nevertheless, it is useful to generate samples that have good "space filling" properties relative to the sample size, particularly when computational burden is not a major consideration.

## 6. Summary and conclusion

In this paper we have extended the analysis in O'Neill (2025) by showing how to implement quantile-stratified (QS) sampling for multivariate distributions. As with the univariate case, QS sampling in the multivariate case involves sampling from a distribution by first breaking the support up into equiprobable quantile blocks and then sampling one value from each block. In the multivariate case the quantile-blocks are formed based on quantiles from a distance measure from the mean/mode of the distribution. We have shown how to generate a QS sample of any size for the multivariate normal distribution, multivariate T-distribution, multivariate logistic distribution, multivariate elliptically contoured distributions, and general strictly quasi-concave continuous multivariate distributions. In each case our simulation method involves taking a scalar distance measure from the mode and using this to generate a univariate QS sample of distances. Once these are obtained, we simulate from the corresponding level curves of the distribution at that distance. For distributions with computational complexity that prevents implementation of this method, we can fall back on quasi-QS sampling by using the distance measure to create an ordering of a set of simulations and drawing regularly spaced order-statistics as a quasi-QS sample.

As noted in O'Neill (2025), the QS sampling method has some advantages over IID sampling for certain problems. The empirical quantiles of the QS samples typically adhere more closely to the true quantiles of the sampling distribution than for an IID sample since the method forces one sample value into each quantile-block of the distribution. Moreover, there is typically less variability in the distances between empirical quantiles in QS samples than in IID samples. This means that QS samples have more stable QQ plots and show a greater adherence to the true quantiles of the sampling distribution. (In multivariate settings, we look at the quantiles based on the distribution of the distance measure.) Generating samples using QS sampling can be useful because it ensures that the generated samples cover each of the quantile-blocks in the distribution. The method is useful in problems involving estimation of mean quantities through simulation from a stipulated sampling distribution, including importance sampling.

Various functions for QS sampling are implemented in the **`utilities`** package in **`R`** (O'Neill 2025). The QS sample for a univariate function is implemented in the **`qs.sample`** function. QS samples for the multivariate normal, multivariate T, and multivariate logistic distributions are implemented in the respective functions **`qs.sample.norm`**, **`qs.sample.t`** and **`qs.sample.logistic`**. Table 1 below shows these functions and their inputs. In each case the user must specify the sample size to generate and the quantile function for the sampling distribution. For the multivariate QS sampling functions the user also inputs the parameters of the distribution at issue. Although not discussed in the present paper, the functions can also generate layered QS samples by adding an input giving the vector of layer sizes for the sample (see O'Neill 2025 for discussion).

**TABLE 1:** Functions for quantile-stratified sampling in the **`utilities`** package

| Function | Inputs |
|---|---|
| `qs.sample` | `n, Q, prob.arg = 'p', layers = NULL, ...` |
| `qs.sample.norm` | `n, mean, var, layers = NULL, ...` |
| `qs.sample.t` | `n, mean, scale, df, layers = NULL, ...` |
| `qs.sample.logistic` | `n, mean, scale, layers = NULL, ...` |
| **Inputs** | **Inputs** |
| `n` | `The number of sample values to be generated (a non-negative integer)` |
| `Q` | `The quantile function of the sampling distribution (must be a function)` |
| `mean` | `The mean vector for the distribution (for norm, t and logistic distributions)` |
| `var` | `The variance matrix for the distribution (for norm distribution)` |
| `scale` | `The scale matrix for the distribution (for t and logistic distributions)` |
| `df` | `The degrees-of-freedom for the distribution (for t distribution)` |
| `prob.arg` | `The name of the probability argument in the quantile function Q` |
| `layers` | `Optional vector giving the number of sample values in each layer of the sample` |
| `...` | `Distribution parameters to be passed through to the quantile function Q` |